\documentstyle[12pt,epsfig]{article}

\def\spose#1{\hbox to 0pt{#1\hss}}
\def\lsim{\mathrel{\spose{\lower 3pt\hbox{$\mathchar"218$}}
 \raise 2.0pt\hbox{$\mathchar"13C$}}}
\def\gsim{\mathrel{\spose{\lower 3pt\hbox{$\mathchar"218$}}
 \raise 2.0pt\hbox{$\mathchar"13E$}}}
 \def\be{\begin{equation}}
 \def\ee{\end{equation}}

\newcommand\npb[3]{Nucl.\ Phys.\ B {\bf #1} (#2) #3}

\newcommand\prd[3]{Phys.\ Rev.\ D {\bf #1} (#2) #3}

\newcommand\rmp[3]{Rev.\ Mod.\ Phys.\ {\bf #1} (#2) #3}

\newcommand\zpc[3]{Z.\ Phys.\ C {\bf #1} (#2) #3}
\newcommand{\hepph}[1]{{\tt hep-ph/#1}}

\begin{document}

\begin{titlepage}

\vspace{0.4cm}
\begin{center}
\Large\bf\boldmath
$B\to\pi \phi$ in SM and MSSM
\unboldmath
\end{center}

\vspace{0.5cm}
\begin{center}
{\sc  Jian-Feng Cheng, Chao-Shang Huang}\\
\vspace{0.7cm}
{\sl  Institute of Theoretical Physics, Academia Sinica.
\\
P.O.Box 2735, Beijing 100080, P.R.China\\}
\vspace{0.3cm}
\end{center}

\vspace{0.3cm}
\begin{abstract}
\vspace{0.2cm}\noindent
We investigate the pure penguin decays $B\to \pi\phi$ in the Standard
Model(SM) and in the Constrained Minimal Supersymmetric Standard Model (CMSSM) using the QCD
factorization approach and consider the Sudakov effects in the twist-3 contribution. We 
find ${\rm Br}(B^-\to \pi^-\phi) = (1.95-5.70)\times 10^{-9}$
in SM and $(1.1-2.4)\times 10^{-8}$ in CMSSM with large 
$\tan\beta $ which is about one order of  magnitude larger than that in SM.
\end{abstract}

\end{titlepage}

\section{Introduction}
One of charmless two-body nonleptonic decays of B mesons, the process $B\rightarrow \phi \pi$, is interesting
because it is a pure penguin process and, in particular, there are no annihilation diagram  contributions
 the importance of which is still in dispute~\cite{li,bbns}. It is sensitive to new physics due to  all
 contributions arising from the penguin diagrams. The calculation of the hadronic matrix element relevant to
 the process is relatively reliable because of no contributions coming from diagrams of 
 annihilation topology. Therefore, we shall investigate the process in both SM and MSSM.
 
 The study of exclusive processes with large momentum transfer in the 
 perturbative QCD (PQCD) has been
extensively carried out and it is shown that the application of PQCD to them is successful~\cite{exclu}.
The key point to apply PQCD is to prove that the factorization, the separation of the short-distance 
dynamics and long-distance dynamics, can be performed for those processes. Recently, two groups, Li et al.
\cite{li1,li} and BBNS~\cite{bbns}, have made  significant progress in calculating hadronic matrix 
elements of local operators relevant to charmless two-body nonleptonic decays of B mesons in the PQCD framework. 
In the letter we shall use BBNS's method ( QCD factorization ) to calculate
the hadronic matrix element of operators relevant to the decay $B\rightarrow \phi \pi$. 

The decay $B\rightarrow \phi \pi$ has been studied by several people~\cite{fle,dg,mel,bey,dun}. 
The naive factorization or BSW model~\cite{bsw} is used in calculating the hadronic matrix 
elements in Refs.\cite{fle,dg}. The modified perturbative QCD approach~\cite{li1}, instead of 
using BSW model, is used in Ref.\cite{mel}. In Ref.\cite{bey}, the QCD improved
factorization (simply, QCD factorization)\cite{bbns} is used and only the leading twist contribution is included.
The numerical result of the branching ratio (Br) in SM given in Ref.\cite{bey} is about an order
 of magnitude larger than that in Refs.\cite{fle,dg}. That is, "non-factorizable" contributions 
(including vertex, penguin, and hard spectator scattering corrections), which are the $O(\alpha_s)$ 
corrections to the leading order result, dominate over the leading order result.
In Ref.\cite{dun} the same QCD factorization is used and the twist-3 contributions are included. However,
the Br given in Ref.\cite{dun} is (3-8)$\times 10^{-10}$ which is of the same order of or even smaller than
that in Refs.\cite{fle,dg}. Considering these disagreements, it is necessary to do a calculation of Br in SM
using the QCD factorization. We carry out such a calculation in SM first and then in Constrained MSSM. The
difference between our calculation and that in \cite{dun} is how to calculate the twist-3 contributions. The 
authors in Ref.\cite{dun} follow BBNS's approach, i.e., to introduce 
a phenomenological parameter instead of the integral containing  
end point singularity~\cite{bbns1}. Indeed, the solution of the end point 
singularity in investigating form factors of mesons is known for a long time~\cite{ff}. That is, to retain
quarks' transverse momenta in both the hard scattering kernels and distribution amplitudes of mesons and to include the Sudakov suppression~\cite{sud} make the integral convergent and computable. Thus, there is no any phenomenological parameter introduced. In the letter we shall use the method to calculate the twist-3 effects. Our numerical result
 of the Br($B^{\pm}\rightarrow \pi\phi$) in SM using QCD factorization
 approach(QCDF) is ${\rm Br}(B^-\to
 \pi^-\phi) = (1.95-5.70)\times 
 10^{-9}$. We have also calculated the Br in Constrained MSSM in order to see supersymmetric (SUSY) effects on the decay.
 The numerical result in Constrained MSSM with large $\tan\beta$ is ${\rm
 Br}(B^-\to \pi^-\phi) =(1.1-2.4)
 \times10^{-8} $ depending on the choice of some relevant parameters.

 The $\Delta B = 1$  effective weak Hamiltonian in SM is given by 

\begin{equation}\label{Heff}
   {\cal H}_{\rm eff} = \frac{G_F}{\sqrt2} \sum_{p=u,c} \!
   \lambda_p \bigg( C_1\,Q_1^p + C_2\,Q_2^p
   + \!\sum_{i=3,\dots, 10}\! C_i\,Q_i + C_{7\gamma}\,Q_{7\gamma}
   + C_{8g}\,Q_{8g} \bigg) + \mbox{h.c.} \,,
\end{equation}
where $\lambda_p=V_{pb}V_{pd}^{*}$, $Q_{1,2}^p$ are the left-handed current--current operators arising
from $W$-boson exchange, $Q_{3,\dots, 6}$ and $Q_{7,\dots, 10}$ are
QCD and electroweak penguin operators, and $Q_{7\gamma}$ and $Q_{8g}$
are the electromagnetic and chromomagnetic dipole operators, respectively. Their explicit expressions can be found in, e.g., Ref.~\cite{bbns1}.
Follow BBNS approach~\cite{bbns},
 the hadronic matrix elements of local operators $Q_i$ at the leading order of the heavy quark expansion can be written as
\begin{eqnarray}\label{fact}
\langle\pi(p)\phi(q)|Q_i|\bar B(p)\rangle&=& F_0^{B\to\pi}(q^2)\int_0^1 dv\, T^I
(v)\Phi_{\phi}(v) \nonumber \\
&&+\int_0^1 d\xi dudv\,
T^{II}(\xi,u,v)\Phi_B(\xi)\Phi_{\pi}(u)\Phi_{\phi}(v) 
\end{eqnarray}
where $\phi_M$ (M=$\phi, \pi, B$) are light-cone distribution amplitudes of 
the meson M, $T^I_{i}$ and $T^{II}_i$ are hard scattering  kernels.
\begin{figure}[t]
\epsfxsize=15cm
\centerline{\epsffile{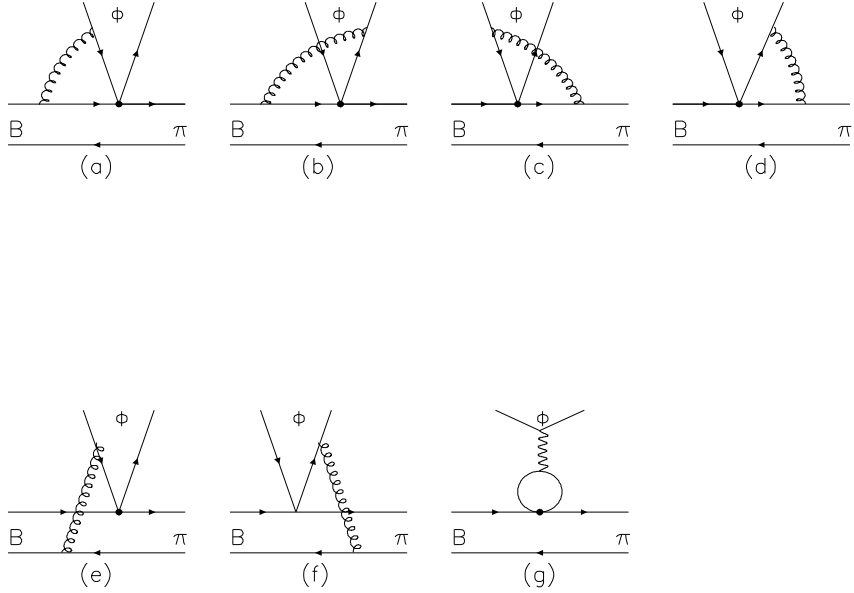}}
\centerline{\parbox{12cm}{\caption{\label{fig1}
Order $\alpha_s$ corrections to the hard-scattering kernels 
$T^{\rm I}$  and $T^{\rm II}$ . 
}}}
\end{figure}

\section{ $\alpha_s$ order corrections of hadronic matrix elements}
The $\alpha_s$ order hard scattering kernels in  Eq.(\ref{fact}) can be
obtained by calculating the diagrams in Fig.(1). Substituting
the kernels into Eq.(\ref{fact}), we get
\be\label{nlome}
 \langle\pi \phi|Q_i|B\rangle_{\alpha_s\, {\rm order}} = \frac{\alpha_s}{4\pi}
 \frac{C_F}{N}F_i  \langle\pi \phi|Q_{i-1}|B\rangle_{{\rm tree}}
 \ee
 where $F_i$=F for i=4,10, (-F-12) for i=6,8 and 0 otherwise  with 
 \begin{eqnarray}\label{g8pi}
F=-12\ln\frac{\mu}{m_b}-18+f^I_\phi+f^{II}_\phi.
\end{eqnarray}
In Eq.(\ref{g8pi}) 
\begin{eqnarray}\label{g8pi1}
&&f^I_\phi=\int^1_0 \!\!\!dx\,g(x)\Phi_\phi(x),\\
&&g(x)=3\frac{1-2x}{1-x}\ln x-i 3 \pi
\end{eqnarray}
is the contribution from the diagrams (a)-(d), the vertex corrections, and $f_{\phi}^{II}$ presents
the contribution from the hard spectator scattering diagrams (e) and (f) which is of real 
non-factorization contribution.
If we ignore the transverse momenta of partons, like that in Eq.(\ref{g8pi1}), 
\be\label{fac}
f_{\phi}^{II}\propto \int d\xi du dv \left[ {\phi_B \left (\xi\right)\over \xi}
                  {\phi_\pi \left(u\right)\over u}
		  {\phi_{||}\left(v\right)\over v}+
 {2 \mu_\pi\over m_B}{\phi_B\left(\xi\right)\over \xi}
                     {{\phi_\sigma\left(u\right)\over 6}\over u^2}
		     {\phi_{||}\left(v\right)\over v} \right],
\ee
where the distribution amplitudes of $\pi, \phi, B$ mesons can be found in Refs.~\cite{bf,vec,ds}.
When $u\to 0$, the twist-3 contribution, the next term of Eq.(\ref{fac}), 
will lead to divergence due to the end-point singularity. Therefore, it is necessary to
consider the transverse momentum $k_T$ effect and include the Sudakov suppression factors
to eliminate the end-point singularity\cite{sud}. After including the transverse momenta of
partons, Eq.(\ref{fac}) changes into
\begin{eqnarray}
f^{II}_\phi &\propto&
\int d\xi du dv\, d^2{\bf k_T} d^2 {\bf k_{1T}} d^2 {\bf k_{2T}}
\nonumber\\
 &\  \times& \Bigg[\ \ {{-um_B^4}
 \phi_B(\xi)\phi_\pi(u)\phi_{||}(v)\over {[\xi
 um_B^2+({\bf k_T }-{\bf k_{1T}})^2][-uvm_B^2+({\bf k_T}-{\bf
 k_{1T}}+{\bf k_{2T}})^2]} }
\nonumber\\ 
&&\ +\
{- 2\mu_{\pi} m_B^5 u v 
 \phi_B(\xi)\frac{\phi_{\sigma}(u)}{6}\phi_{||}(v)\over [\xi
 um_B^2+({\bf k_T }-{\bf k_{1T}}^2)][-uvm_B^2+({\bf k_T}-{\bf
 k_{1T}}+{\bf k_{2T}})^2]^2} \Bigg] \,.
\end{eqnarray} 

The $k_T$ resummation of
large logarithmic corrections to $B$, $\phi$ and $\pi$ meson distribution
amplitudes leads to the presence of the exponentials $S_B$,$S_\phi$ and $S_\pi$
respectively\cite{li2},
\begin{eqnarray}
S_{B}(t)&=&\exp\left[-s(\xi P_{B}^{+},b)
-2\int_{1/b}^{t}\frac{d{\bar{\mu}}} {\bar{\mu}}
\gamma (\alpha _{s}({\bar{\mu}}^2))\right]\;,
\nonumber \\
S_{\phi }(t)&=&\exp\left[-s(vP_{\phi}^{+},b_{2})
-s((1-v)P_{\phi}^{+},b_{2})
-2\int_{1/b_{2}}^{t}\frac{d{\bar{\mu}}}{\bar{\mu}}
\gamma (\alpha _{s}({\bar{\mu}}^2))\right]\;,
\nonumber \\
S_{\pi}(t)&=&\exp\left[-s(uP_\pi^{-},b_{1})
-s((1-u)P_{\pi}^{-},b_{1})
-2\int_{1/b_{1}}^{t}\frac{d{\bar{\mu}}}{\bar{\mu}}
\gamma (\alpha_{s}({\bar{\mu}}^2))\right]\;,
\label{sbk}
\end{eqnarray}
with the quark anomalous dimension $\gamma=-\alpha_s/\pi$.
The variables $b$, $b_{1}$, and $b_{2}$, corresponding to the parton
transverse momenta $k_T$, $k_{1T}$, and $k_{2T}$ respectively, represent the
transverse extents of the $B$, $\pi$ and $\phi$ mesons, respectively.
The expression for the exponent $s$ is referred to \cite{CS,BS,LS}.
The above Sudakov exponentials decrease so fast in the large $b$ region
that the $B\to\pi \phi$ hard amplitudes remain sufficiently perturbative
in the end-point region. 
By a straightforward calculating, we obtain
the hard spectator scattering contribution 
\begin{eqnarray}\label{hbpi}
f^{II}_{\phi}&=&\frac{4\pi^2}{N}\frac{f_\pi f_B}{F^{B\rightarrow\pi}_+(m^2_\phi) m^2_B}
\int d\xi dudv
\int bdb b_2db_2 \nonumber\\
& \times&
\Bigg\{- u m_B^4 {\cal P}_B(\xi,b){\cal P}_{\pi}(u,b)
{\cal P}_{||}(v,b_2)K_0(-i\sqrt{uv}m_B b_2)\nonumber \\
&&\ \ \ \times \left[\theta(b_2-b)I_0(\sqrt{\xi
u}m_Bb)K_0(\sqrt{\xi u}m_B b_2)  +
\theta(b-b_2)I_0(\sqrt{\xi u}m_Bb_2)K_0(\sqrt{\xi u}m_B b)
\right]\nonumber\\
&& -2uv\mu_{\pi}m_B^5 {\cal P}_B(\xi,b)
\frac{{\cal P}_{\sigma}(u,b)}{6}{\cal P}_{||}(v,b_2)
 \frac{b_2}{-2i\sqrt{uv}m_B}K_{-1}(-i\sqrt{uv}m_B b_2)\nonumber\\
 &&\ \ \ 
\times\left[\theta(b_2-b)I_0(\sqrt{\xi u}m_Bb)K_0(\sqrt{\xi u}m_B b_2)+
 \theta(b-b_2)I_0(\sqrt{\xi
u}m_Bb_2)K_0(\sqrt{\xi u}m_B b)  \right] ~\Bigg\}\nonumber\\
\ 
\end{eqnarray} 

where $f_\pi$ ($f_B$) is the pion ($B$) meson decay constant,
$m_B$ the $B$ meson mass, $F_+^{B\rightarrow\pi}(m_\phi^2)$ the $B\to\pi$ 
form factor
at the momentum transfer $m^2_\phi$, $\xi$ the light-cone momentum fraction
of the spectator in the $B$ meson, $K_i$, $I_i$ are modified Bessel functions of order $i$ and ${\cal
P}_B$, ${\cal P}_\pi$, ${\cal P}_\phi$ are $B$, $\pi$, $\phi$ corrected meson 
amplitudes with the exponentials $S_B$, $S_\phi$ and $S_\pi$ respectively~\cite{li2}. As noted in Refs.\cite{ff},
although the twist-3 contribution is power suppressed it is numerically comparable with the twist-2 contribution
due to the chirally-enhanced factor $m_{\pi}^2(\mu)/m_B(\mu)[\bar{m}_u(\mu)+\bar{m}_{d}(\mu)]$. 
From Eq.(\ref{hbpi}), we obtain that the twist-3 contribution to $f_{II}$ is numerically about
the fourth of the twist-2 contribution to $f_{II}$. It is worth to note that the numerical
result of the twist-2 contribution to $f_{II}$ obtained from Eq. (\ref{hbpi}) is  almost 
completely the same as that obtained without including the Sudakov factor, as expected.

\section{The branching ratio in SM}
The effective Hamiltonian (\ref{Heff}) results in the following matrix element for the decay
\begin{equation}\label{Top}
   \langle\pi \phi|{\cal H}_{\rm eff}| B\rangle
   = \frac{G_F}{\sqrt2} \sum_{p=u,c} \lambda_p\,
   \langle\pi \phi|{\cal T}_p| B\rangle \,,
\end{equation}
where 
\begin{eqnarray}\label{Toper}
   {\cal T}_p &=&
   a_3(\pi \phi) \, (\bar d b)_{V-A} \otimes
    (\bar s s)_{V-A} \nonumber\\
   &+& a_5(\pi \phi) \, (\bar d b)_{V-A} \otimes
    (\bar s s)_{V+A} \nonumber\\
   &+& a_7(\pi \phi) \, (\bar d b)_{V-A} \otimes
    {\textstyle\frac32} e_s(\bar s s)_{V+A} \nonumber\\
   &+& a_9(\pi \phi)\, (\bar d b)_{V-A} \otimes
    {\textstyle\frac32} e_s(\bar s s)_{V-A}\,. \nonumber\\
\end{eqnarray}
The symbol $\otimes$ indicates that 
the matrix elements of the operators in ${\cal T}_p$ are to be evaluated 
in the factorized form $\langle\pi \phi|j_1\otimes j_2| B\rangle\equiv
\langle\pi|j_1| B\rangle\,\langle \phi|j_2|0\rangle$. The $O(\alpha_s)$ corrections, including the
nonfactorizable corrections corresponding to the diagrams (e) and (f) in Fig.1, of hadronic matrix elements are, by definition, included in the coefficients $a_i$. Collecting the results in the above section, we have
\begin{eqnarray} 
a_3(\pi\phi) &=& C_3+\frac{1}{N}C_4 +
\frac{\alpha_s}{4\pi}\frac{C_F}{N}\,C_4\,F, 
\label{a3}\\
a_5(\pi\phi) &=& C_5+\frac{1}{N}C_6 +
\frac{\alpha_s}{4\pi}\frac{C_F}{N}\,C_6\,(-F-12), 
\label{a5}\\
a^p_7(\pi\phi)&=& C_7+{C_8\over N}+{\alpha_s\over
4\pi}\frac{C_F}{N}\,C_8\,(-F-12)+ {\alpha_{\rm em}\over 9 \pi}
P^p_{\rm em} (C_1+3 C_2), \label{a7} \\
a^p_9(\pi\phi) &=& C_9+\frac{1}{N}C_{10} +
\frac{\alpha_s}{4\pi}\frac{C_F}{N}\,C_{10}\,F + {\alpha_{\rm em }\over 9 \pi
}P^p_{\rm em} (C_1+3 C_2)\label{a9}\,,
\end{eqnarray}
which has form of $A+B \alpha_s$. In order to keep our calculation consistent, we 
use LO Wilson coefficients which contribute to B and NLO Wilson
coefficients which contribute to A. 
In Eqs. (\ref{a3}-\ref{a9}) $C_F=(N^2-1)/(2N)$ ( $N=3$  is the number
of colors ), and $P^p_{\rm em}$ arises from electroweak penguin
contributions, Fig. \ref{fig1}(g), and is given by
\begin{eqnarray}
P^p_{\rm em}&=& {10\over 9}-4 \int^1_0 du u (1-u) \ln \left({m^2_q-q^2 u
(1-u)\over \mu^2}\right).
\end{eqnarray}

From Eq.(\ref{Top}), the decay amplitude for $B^-\to \pi^-\phi$ is 
\begin{eqnarray}\label{amp}
&&A(B^-\to \pi^-\phi)\nonumber\\
&=&\sqrt{2}\,A(B^0\to \pi^0\phi)\nonumber \\
&=& {G_F\over\sqrt{2}}\sum\limits_{p=u,c}\lambda_p \left[ a_3+a_5
-{1\over 2 }\left(a_7^p+a_9^p\right)\right]\,
f_\phi m_\phi 
F^{B\to\pi}_+(m^2_\phi)~2~\epsilon^\phi_L\cdot p_B ,
\end{eqnarray}
 The relevant Wilson coefficients in NDR
scheme are
showed in Table \ref{t1}~\cite{bbns}.
\begin{table}[t]  
\centerline{\parbox{14cm}{\caption{\label{t1}
Wilson coefficients $C_i$ in the NDR scheme. Input parameters are 
$\Lambda^{(5)}_{\overline{\rm MS}}=0.225$\,GeV, $m_t(m_t)=167$\,GeV, 
$m_b(m_b)=4.2$\,GeV, $M_W=80.4$\,GeV, $\alpha=1/129$, and 
$\sin^2\!\theta_W=0.23$.}}}
\begin{center}
\begin{tabular}{|l|c|c|c|c|c|c|}
\hline\hline
NLO & $C_1$ & $C_2$ & $C_3$ & $C_4$ & $C_5$ & $C_6$ \\
\hline
$\mu=m_b/2$ & 1.137 & $-0.295$ & 0.021 & $-0.051$ & 0.010 & $-0.065$ \\
$\mu=m_b$   & 1.081 & $-0.190$ & 0.014 & $-0.036$ & 0.009 & $-0.042$ \\
$\mu=2 m_b$ & 1.045 & $-0.113$ & 0.009 & $-0.025$ & 0.007 & $-0.027$ \\
\hline
 & $C_7/\alpha$ & $C_8/\alpha$ & $C_9/\alpha$ & $C_{10}/\alpha$
 & $C_{7\gamma}^{\rm eff}$ & $C_{8g}^{\rm eff}$ \\
\hline
$\mu=m_b/2$ & $-0.024$           & 0.096 & $-1.325$ & 0.331
 & --- & --- \\
$\mu=m_b$   & $-0.011$           & 0.060 & $-1.254$ & 0.223
 & --- & --- \\
$\mu=2 m_b$ & $\phantom{-}0.011$ & 0.039 & $-1.195$ & 0.144
 & --- & --- \\
\hline\hline
LO & $C_1$ & $C_2$ & $C_3$ & $C_4$ & $C_5$ & $C_6$ \\
\hline
$\mu=m_b/2$ & 1.185 & $-0.387$ & 0.018 & $-0.038$ & 0.010 & $-0.053$ \\
$\mu=m_b$   & 1.117 & $-0.268$ & 0.012 & $-0.027$ & 0.008 & $-0.034$ \\
$\mu=2 m_b$ & 1.074 & $-0.181$ & 0.008 & $-0.019$ & 0.006 & $-0.022$ \\
\hline
 & $C_7/\alpha$ & $C_8/\alpha$ & $C_9/\alpha$ & $C_{10}/\alpha$
 & $C_{7\gamma}^{\rm eff}$ & $C_{8g}^{\rm eff}$ \\
\hline
$\mu=m_b/2$ & $-0.012$           & 0.045 & $-1.358$ & 0.418
 & $-0.364$ & $-0.169$ \\
$\mu=m_b$   & $-0.001$           & 0.029 & $-1.276$ & 0.288
 & $-0.318$ & $-0.151$ \\
$\mu=2 m_b$ & $\phantom{-}0.018$ & 0.019 & $-1.212$ & 0.193
 & $-0.281$ & $-0.136$ \\
\hline\hline
\end{tabular}
\end{center}
\end{table}
For the other parameters, we use
 \begin{eqnarray}
&&f_B=0.190\,{\rm GeV}, \hspace{0.5cm} f_\pi = 0.131 {\ \rm Gev}, \nonumber\\ 
&&f_\phi = 0.237 {\ \rm Gev},\hspace{0.5cm}
f_\phi^T=0.215{\rm \ Gev},\hspace{0.5cm} F^{B\to\pi}_+(m^2_\phi)=0.30\,
\end{eqnarray}
and Wolfensein parameters fitted by Ciuchini as \cite{ciuchini} 
\begin{eqnarray}
&&A= 0.819\,,\hspace{1.0cm} \lambda=0.224\,, \hspace{1.0cm} \nonumber \\
&&\bar{\rho}=\rho (1- \lambda^2/2)=0.224\,,\hspace{0.5cm}\bar{\eta}=\eta
(1-\lambda^2/2)=0.317\,.
\end{eqnarray}
The numerical results of the branch
ratio at different scales are showed 
in Table {\ref{br}.
\begin{table}[t] 
\centerline{\parbox{14cm}{\caption{\label{br}
Branch ratios of $B^-\to \pi^-\phi $ at scale $m_b/2$, $m_b$ and $2\,m_b$.
}}}
\begin{center}
\begin{tabular}{l|c|c|c|c|c|c}
\hline\hline
&\multicolumn{2}{|c|}{$\mu=m_b/2$}&\multicolumn{2}{|c|}{$\mu=m_b$}
&\multicolumn{2}{|c}{$\mu=2\,m_b$} \\
\hline
& NF & QCDF & NF & QCDF & NF & QCDF \\
\hline
\hline
Br in SM
& $2.47\times 10^{-9}$ &$ 1.95\times
10^{-9}$&$ 6.53\times 10^{-10}$&$ 4.51\times 10^{-9}$&$ 2.52\times 10^{-9}$&$
5.70\times 10^{-9}$\\
Br in CMSSM &$  $&$2.4\times10^{-8}$&$  $&$
1.6\times10^{-8}$&$  $&$1.1\times10^{-8}$\\
\hline
\hline
\end{tabular}
\end{center}
\end{table}
\section{Br in Constrained MSSM}
It has been shown that the neutral Higgs bosons (NHBs) do make significant contributions to leptonic and
semileptonic rare B decays in Constrained MSSM with large $\tan\beta$~\cite{hy,huang,urban}. For $b\rightarrow
ds\bar{s}$, it is expected that the similar enhancement of Br will happen since the mass of the strange 
quark is the same order as or a little of larger than that of muon. In the section we will calculate the Br of $B\rightarrow
\pi\phi$ in the large $\tan\beta$ case of the Constrained MSSM.

In addition to Eq.(\ref{Heff}), we have ~\cite{dhh,hy}
\begin{equation}\label{Heffn}
   {\cal H}_{\rm eff}^{new} = \frac{G_F}{\sqrt2}(-\lambda_t) 
    \!\sum_{i=11,\dots, 16}\! C_i\,Q_i 
   + \mbox{h.c.} \,,
\end{equation}
where $Q_{11}$ to $Q_{16}$, the neutral Higgs penguins operators, are given by
\begin{eqnarray}
Q_{11} &=& (\bar d b)_{S+P} \sum{}_{\!q}\,(\bar q q)_{S-P} \,, \hspace{1.2cm}
Q_{12} = (\bar d_i b_j)_{S+P} \sum{}_{\!q}\,(\bar q_j q_i)_{S-P} \,, \nonumber\\
Q_{13} &=& (\bar d b)_{S+P} \sum{}_{\!q}\,(\bar q q)_{S+P} \,,
    \hspace{1.2cm}
Q_{14} = (\bar d_i b_j)_{S+P} \sum{}_{\!q}\,(\bar q_j q_i)_{S+P} \,, \nonumber\\
Q_{15} &=& \bar d \sigma^{\mu\nu}(1+\gamma_5) b \sum{}_{\!q}\,
    \bar q \sigma_{\mu\nu}(1+\gamma_5)q \,, \nonumber\\
Q_{16} &=& \bar d_i \sigma^{\mu\nu}(1+\gamma_5) b_j \sum{}_{\!q}\,
    \bar q_j \sigma_{\mu\nu}(1+\gamma_5) q_i \,.
\end{eqnarray}
Here $(\bar q_1 q_2)_{S\pm P}=\bar q_1(1\pm\gamma_5)q_2$. Then Eq.(\ref{Top}) is extended to
\begin{equation}\label{Topn}
   \langle\pi \phi|{\cal H}_{\rm eff}| B\rangle
   = \frac{G_F}{\sqrt2} \sum_{p=u,c} \lambda_p\,
   \langle\pi \phi|{\cal T}_p+{\cal T}_p^{\rm neu}| B\rangle \,,
\end{equation}
where the term ${\cal T}_p^{neu}$ arises from the neutral Higgs
  contributions, given by
\begin{eqnarray}\label{neu}
   {\cal T}^{neu}_p &=& r_\chi^\phi(\mu)\,\big[a_{11}(\pi \phi)\,
   + a_{13}(\pi \phi) \big]
    (\bar d b)_{V+A} \otimes (\bar s s)_{V-A}\,. 
\end{eqnarray}
The Wilson coefficients $C_{Q_i}$ (i=11, ..., 16) in Eq.(\ref{Heffn}) is calculated in the same way as that in 
Ref.\cite{hy,huang,urban} and results are
\begin{eqnarray}
 C_{Q_{11}}(M_W) &= &
-{\alpha_{\rm em}\over 4\pi}\,
\frac{m_b m_s\,\tan^3\beta}{4 \sin^2\theta_w M_W^2 \lambda_t}
\sum_{i=1}^2 \sum_{k=1}^6 U_{i2}
T^{km}_{UL} K_{m b} \left[-\sqrt{2} V_{i1} (T_{UL} K)^*_{ks}+ V_{i2} 
\frac{ (T_{UR} 
{\tilde m_u} K)^*_{ks}}{M_W \sin\beta} \right]\nonumber \\
&&\times\ (r_{hH}+r_A) \sqrt{x_{\chi_i^-}}
f_{B^0}\left(x_{\chi_i^-},
x_{\tilde u_k}\right) +O(\tan^2\beta),\label{c1a}\\
C_{Q_{13}}(M_W) &= & -{\alpha_{\rm em}over 4\pi }
 \frac{m_b m_s\, \tan^3\beta }{4 \sin^2\theta_w M_W^2 \lambda_t}
 \sum_{i=1}^2 \sum_{k=1}^6 U_{i2}
T^{km}_{UL} K_{m b} \left[-\sqrt{2} V_{i1} (T_{UL} K)^*_{ks}+ V_{i2} \frac{ (T_{UR} 
{\tilde m_u} K)^*_{ks}}{M_W \sin\beta} \right]\nonumber \\
&&\times\ (r_{hH}-r_A) \sqrt{x_{\chi_i^-}}
f_{B^0}\left(x_{\chi_i^-},
x_{\tilde u_k}\right) +O(\tan^2\beta),\label{c1a}\\
 C_{Q_i}(m_W)&=&0, ~~~~~~~i=12, 14, 15, 16,
\end{eqnarray}
where the definitions of various symbols are the same as those in
Ref.\cite{huang}.
In calculating $C_{Q_i}(m_W)$ the following values of relevant parameters 
are used\footnote{In CMSSM the mass spectrum and mixing of sparticles and Higgs bosons
 can be calculated given a set of values of a few parameters at the high (GUT or Planck) scale. Here, in stead of scanning the parameter space, we 
take reasonable values of relevant parameters for the sake of simplicity.}:
\begin{eqnarray}\label{para}
&&\tan\beta=60,\ \ \ m_{h^0}=110\ {\rm Gev},\ \ \ m_{H^0}=150\
{\rm Gev},
\nonumber\\
&&m_{H^-}=200{\rm Gev},\ \ \ m_b=4.2\ {\rm Gev},\ \ \ M_2=320\ {\rm Gev},\ \
\ \mu=270{\ {\rm Gev}}, \nonumber\\
&&m_{t_1}=120\ {\rm Gev},\ \ \ \theta_{\tilde t_1}=-\pi/4,\ \ \
m_s=0.11\,{\rm Gev}.
\end{eqnarray}
At the low scale we get 
\begin{eqnarray}
C_{Q_{12}}+C_{Q_{14}}= 0.139, 0.0897, 0.0565\ {\rm for}\  \mu=m_b/2, m_b, 2m_b.
\end{eqnarray}
$a_i$ in Eq. (\ref{neu}) in MSSM read as 
\begin{eqnarray}
&&a_{11,13} = -{\alpha_s\over 4\pi } \, {C_F\over N}\, (f^I_s +  f^{II}_s ) \,
C_{Q_{12,14}}\,, \\
\end{eqnarray}
where 
\begin{eqnarray}
&&f^I_s = - 2\,\int ^1 _0 du  \big[ \ln^2 u + 2\ln u - 2{\rm Li}_2 (u) \big]
\, \phi_s(u)\nonumber\\ 
&&\hspace{0.4cm} + 2 \int^1 _0 du \int^{m_b}_0 dk \int db \,
\ln \left[ \sqrt{ {4k^2_T\over m_b^2 } +u^2 } +u \right] \, J_0 ( bk )
{\cal P}_s ( u,b) \\
&& f^{II}_s =   { 2\pi\,m_B \, f_\pi f_B\over F^{B\to \pi}_+{m^2_\phi}} 
\int [du][db]\,
\delta ^2 (b_1 +b_2 )\, b\, {\cal P}_B(\xi,b)\, {\cal P}_s(v,b_2)\nonumber\\ 
&&\hspace{0.4cm}\times \big[ \mu_p\,
(u+v)\,{\cal P}_p(u,b_1)+m_B\,(\xi-v) \,{\cal P}(u,b_1)\big] \nonumber\\ 
&&\hspace{0.4cm} \times \left[ \theta(b_2-b)I_0(b\sqrt{u\xi}\,m_
B)\,K_0(b\sqrt{u\xi} \, m_B) +\theta(b-b_2 )\, I_0 (b\sqrt{u\xi}\,m_B )\,
K_0 (\sqrt{-uv}m_b b_2 ) \right]\,. \nonumber\\ 
\ \ 
\end{eqnarray}
We found in numerical results that the last terms of the distribution amplitude $\phi_s(u)$ of
$\phi$ meson make main contributions to $a_{11}$. 
 
The amplitude for for $B^-\to \pi^-\phi$ now is given as 
\begin{eqnarray}
&&A(B^-\to \pi^-\phi)=\nonumber\\
&&{G_F\over\sqrt{2}}\sum\limits_{p=u,c}\lambda_p \left[ a_3+a_5
-{1\over 2 }\left(a_7^p+a_9^p\right)+ r_\chi^\phi\,(a_{11}+a_{13})\right]
\times f_\phi m_\phi 
F^{B\to\pi}_+(m^2_\phi)~2~\epsilon^\phi_L\cdot p_B\, ,
\end{eqnarray}
where 
\begin{equation}\label{rKdef}
   r_\chi^\phi(\mu)
   =  {m_B\over 4 \epsilon \cdot p_B}\, {f_\phi^T \over f_\phi} \,.
\end{equation}
In calculating $a_i$ (i=3,5,7,9) we should use the relevant Wilson coefficients in CMSSM. However,  we still use their SM values in numerical calculations for simplicity
 because the contributions of SUSY only modify
them in a few percents in most part of the parameter space including the values given above, Eq. (\ref{para})~\cite{hyn}.
 The numerical result of the Br is shown in Table \ref{br}.
\section{Summary}
 In summary, we have studied the pure penguin process $B^-\to \pi^-\phi$
using QCD factorization approach, in particular, calculated the twist-3 contribution by including the Sudakov
effects. We find ${\rm Br}(B^-\to\pi^-\phi)=(1.95-5.70)\times10^{-9}$ in SM, which is roughly in agreement with that in Ref.~\cite{mel}.
 Comparing with the naive factorization (NF) result which is $(0.7-3)\times 10^{-9}$, 
the QCD factorization result ( to the $O(\alpha_s)$ )
is less sensitive to the decay scale, as can be seen from Table \ref{br}\footnote{In NF case
the Br at $\mu=2~m_b$ is close to that at $\mu=m_b/2$ while the amplitude at $\mu=2~m_b$ is 
close to that at $\mu=m_b/2$ in magnitude but with the opposite sign.}.
Actually, as noticed in Ref.\cite{chenghy}, the coefficients $a_i$ given in 
Eq.(\ref{a3}-\ref{a9}) are scale independent to the $O(\alpha_s)$, 
which can be demonstrated
by using the LO anomalous dimension matrix of relevant operators. However, from Table \ref{br}
one can see that there still is the significant scale dependence. 
  Because we use the NLO Wilson coefficients
the significant scale dependence comes mainly from the $O(\alpha_s)$ 
corrections of hadronic matrix 
elements. Indeed the $O(\alpha_s)$ corrections of hadronic 
matrix elements depend heavily on
the scale (see, Eq.(\ref{g8pi}), which contains the 
factor $12\ln [{\mu}/{m_b}]$). In order to 
decrease the scale 
dependence it is expected to calculate the $\alpha_s^2$ order corrections of 
hadronic matrix elements. We have also
calculated the Br in Constrained MSSM with large $\tan \beta$ and the result is
 ${\rm Br}(B^-\to\pi^- \phi)=(1.1-2.4)\times10^{-8}$. That is, the Br 
 can be enhanced by one order of magnitude  at most 
 compared with that in SM, which is still far below the  Babar experimental bound $B_r(B^{0,\pm}\rightarrow
 \pi^{0,\pm}~\phi)< 5.6\times 10^{-7}$ at $90\%$
 CL~\cite{babar}.

\section*{Acknowledgment}
The work was supported in part by the National Nature Science Foundation of China. One of the authors
(Chao-Shang Huang) would like to thank C.S. Lam for discussions and Phys. Dept., McGill University
where the paper was written for the warm hospitality.

\end{document}